\documentclass[aps,prl,twocolumn,showpacs]{revtex4}
\usepackage{graphicx}
\usepackage{natbib}

\begin{document}

\title{Origin of non-exponential relaxation in a crystalline ionic conductor: a multi-dimensional
$^{109}$Ag NMR study}

\author{M.\ Vogel}
\author{C.\ Brinkmann}
\author{ H.\ Eckert}
\author{A.\ Heuer}
\affiliation{Institut f\"ur Physikalische Chemie, 
Westf\"alische Wilhelms-Universit\"at, Schlossplatz 4/7,
48149 M\"unster, Germany \\ and Sonderforschungsbereich 458}

\date{\today}

\begin{abstract}

The origin of the non-exponential relaxation of silver ions in the
crystalline ion conductor $\mathrm{Ag_7P_3S_{11}}$ is analyzed by
comparing appropriate two-time and three-time $^{109}$Ag NMR
correlation functions. The non-exponentiality is due to a rate
distribution, i.e., dynamic heterogeneities, rather than to an
intrinsic non-exponentiality. Thus, the data give no evidence for
the relevance of correlated back-and-forth jumps on the timescale
of the silver relaxation.

\end{abstract}
\pacs{64.70}
\maketitle

Solid ion conductors are interesting materials from the viewpoints of
both fundamental and applied science. While fast ion transport has
important applications in energy and information technologies, many
of its dynamic aspects are poorly understood at the atomic level. A
prominent feature of many solid ion conductors is their
non-exponential relaxation behavior, reflecting complex ion dynamics
\cite{Moynihan,Angell,Sidebottom}. Two fundamentally different
explanations are possible \cite{12}. In the heterogeneous scenario,
all particles are random-walkers, but a distribution of correlation
times $G(\lg \tau)$ exists. In the homogeneous scenario, all
particles obey the same relaxation function, which is, however,
intrinsically non-exponential due to correlated back-and-forth jumps.
For solid ion conductors, the concept of correlated forward-backward
motion is one of the cornerstones of current modelling approaches. It
is suggested by a strong frequency dependence of the electric
conductivity $\sigma(\nu)$ \cite{Jonscher,Ingram,Roling}. Thus, it
might be tempting to conclude that the homogeneous scenario applies
to ion dynamics, i.e., ion transport might be slowed down by extended
back-and-forth jump sequences.

Despite the technological importance of solid ion conductors, the
origin of the non-exponential relaxation has remained unresolved to
the present date. This is due to the fact that two-time correlation
functions (2T-CF) such as scattering functions or conductivities are
intrinsically unable to decide between both scenarios, but rather
three-time correlation functions (3T-CF) are required \cite{12}. The
lifetimes of dynamic heterogeneities, if they exist, can be measured
by four-time correlation functions (4T-CF). As was shown for
applications in supercooled liquids \cite{12,Spiess,Roessler}, all of
these multi-time correlation functions (MT-CF) are available from
multi-dimensional NMR experiments. There, the 3T-CF has a simple
interpretation: a dynamic filter meant to select slow particles is
applied in the first time interval and the relaxation of the selected
subset is observed in the second time interval. If dynamic
heterogeneities exist, such a selection is possible and the 3T-CF
decays more slowly than the regular 2T-CF.

For solid ion conductors, previous 2T-CF NMR studies have confirmed
that relaxation is non-exponential
\cite{Bohmer,Vogel_JNCS,Vogel_PCCP, Bohmer_2}. Moreover, it has been
shown that MT-CF monitoring the diffusion of single silver ions can
be recorded in $^{109}$Ag NMR \cite{Vogel_PCCP}. In this letter, we
use multi-dimensional $^{109}$Ag NMR to study for the first time the
nature of the non-exponential relaxation in a solid ion conductor in
detail. Specifically, we measure 3T-CF in polycrystalline
$\mathrm{Ag_7P_3S_{11}}$ to resolve the homogeneous and the
heterogeneous contributions to the non-exponential silver ionic
relaxation. Moreover, the lifetime of the dynamic heterogeneities is
quantified by means of 4T-CF. $\gamma$-$\mathrm{Ag_7P_3S_{11}}$
exhibits a high dc conductivity
$\sigma_{dc}\!=\!2.5\times10^{-4}\,\mathrm{S/cm}$ at
$T\!\approx\!300\mathrm K$ \cite{Zhang}. We study silver dynamics in
the $\beta$-phase below a phase transition at $T\!=\!209\mathrm K$
\cite{Brinkmann}. To improve the signal-to-noise ratio
$\mathrm{Ag_7P_3S_{11}}$ was synthesized starting from $^{109}$Ag
enriched (97\%) metal powder. Further experimental details are
described in Refs.\ \cite{Vogel_JNCS,Vogel_PCCP}.

In solid-state $^{109}$Ag NMR, the chemical shift (CS) interaction
dominates the observed frequency shift, $\omega$. The CS tensor
describes the magnetic shielding of the applied static magnetic field
$\mathbf{B_0}$ at the nuclear site due to neighboring electrons. In
our case, fast silver rattling motions in the potential minimum
affect the CS tensor so that a single, time-independent $\omega$ can
be ascribed to each silver site. While the different local
environments result in distinguishable values of $\omega$, the NMR
frequencies at sites related by translational symmetry are identical.
Thus, any change in the resonance frequency is due to silver jumps
between distinguishable sites. In other words, the sites within the
unit cell provide a small number of NMR frequencies which are
discontinuously adopted by the silver ions. $^{109}$Ag NMR MT-CF
directly probe the time dependence of $\omega$ and, hence, silver
diffusion on the ms-s timescale. Since the values of $\omega$
associated with the silver sites depend on the orientation of the
crystal with respect to $\mathbf{B_0}$ a powder average is observed
for our polycrystalline sample.

The measurement of NMR MT-CF is well described in the literature
\cite{Spiess,Roessler}. In general, suitable multi-pulse sequences
are applied to manipulate the spin system. The pulses of these
sequences divide the experimental time into a series of alternating
short evolution times $t_p\!\ll\!\tau$ during which the respective
NMR frequencies are detected and long mixing times
$t_m\!\approx\!\tau$ during which dynamics may take place. The
lengths and the phases of the pulses depend on the spin $I$ of the
observed nucleus. In our case $I\!=\!1/2$, the multi-pulse sequences
can be adopted from $^{13}$C NMR experiments
\cite{Spiess,Roessler,Tracht}. In detail, the three-pulse sequence
$P_1-t_p-P_2-t_m-P_3-t_p$, or stimulated-echo sequence, is used to
record $^{109}$Ag NMR 2T-CF. Varying $t_m$ for constant $t_p$ we
measure
\begin{eqnarray}
&F_2^{cc}(t_m)\propto\,\langle\,\cos[\omega_1 t_p]\cos[\omega_2 t_p]
\,\rangle \:\:\:
\mathrm{and}&\nonumber\\
&F_2^{ss}(t_m)\propto \langle\,\sin[\omega_1 t_p]\sin[\omega_2 t_p]
\,\rangle . &\nonumber
\end{eqnarray}
Here, $\omega_1$ and $\omega_2$ are the frequencies during the two
evolution times separated by the mixing time $t_m$. The brackets
$\langle\dots\rangle$ denote ensemble averages. Suitable seven-pulse
sequences can be applied to correlate the frequencies $\omega_1$,
$\omega_2$, $\omega_3$ and $\omega_4$ during four evolution times
$t_p$ separated by three mixing times $t_{m1}$, $t_{m2}$ and $t_{m3}$
\cite{Spiess,Roessler,Tracht}. We record the 4T-CF
\begin{displaymath}
F_4(t_{m1},t_{m2},t_{m3})\!\propto\!\langle\,\cos[(\omega_2\!-\!\omega_1)t_p\,]\cos
(\omega_3t_p)\cos(\omega_4t_p)\,\rangle,
\end{displaymath}
which has proven very useful for a study of the lifetime of dynamic
heterogeneities. A 3T-CF suited to analyze the origin of the
non-exponential relaxation is obtained for $t_{m2}\!\rightarrow\!0$
in the seven-pulse sequences, i.e., $\omega_3\!=\!\omega_2$:
\begin{displaymath}
F_3(t_{m1},t_{m3})\!\propto\!\langle\,\cos[(\omega_2\!-\!\omega_1)t_p\,]\cos(\omega_2t_p)\cos
(\omega_4t_p)\,\rangle.\nonumber
\end{displaymath}

A straightforward interpretation of these MT-CF is possible when
\emph{each} silver jump to a distinguishable site leads to a complete
loss of correlation. This condition can be met when large evolution
times are applied so that $\Delta \omega t_p\!\gg\!2\pi$ is fulfilled
for basically all frequency changes resulting from these jumps. For
$\Delta \omega t_p\!\gg\!2\pi$, it is easily seen using the
trigonometric addition theorems that
\begin{equation}\label{F2}
F_2(t_m)\equiv F_2^{cc}(t_m)= F_2^{ss}(t_m) \propto
\langle\,\cos[(\omega_2\!-\!\omega_1)t_p]\,\rangle
\end{equation}
where the term $\cos[(\omega_2\!-\!\omega_1)t_p]$ approximates the
$\delta$ function $\delta(\omega_2\!-\!\omega_1)$
\cite{Spiess,Roessler}. In $^{109}$Ag NMR studies of
Ag$_7$P$_3$S$_{11}$, Eq.\ \ref{F2} is fulfilled for
$t_p\!\geq\!100\mu s$ \cite{Vogel_JNCS}. For such evolution times, 
independent of the actual value of $t_p$, $F_2(t_m)$ quantifies the
fraction of ions that occupy the same or -- due to the identical
$\omega$ -- any periodic site after a time $t_m$. Similarly, one can 
rewrite
\begin{eqnarray}
F_3(t_{m1}\!,t_{m3})\!\propto \!\langle\,\cos[(\omega_2\!-\!\omega_1)t_p\,]\cos[(\omega_4\!-\!\omega_2)t_p\,]\,\rangle \:\:\:\mathrm{and}\:\:\label{F3}\\
F_4(t_{m1}\!,t_{m2},t_{m3})\!\propto\!\langle\,\cos[(\omega_2\!-\!\omega_1)t_p\,]\cos[(\omega_4\!-\!\omega_3)t_p\,]\,\rangle.\:\:\:\label{F4}
\end{eqnarray}
In studies of glassy ion conductors, i.e., in the absence of periodic
sites, the term $\cos[(\omega_2\!-\!\omega_1)t_p]$ acts as a perfect
dynamic filter which selects immobile ions during $t_{m1}$
($\omega_2\!=\!\omega_1$). Then, the relaxation of these selected
ions is observed via $F_3(t_{m3})$ so that
$F_3(t_{m3})\!\neq\!F_2(t_{m3})$ directly indicates the existence of
dynamic heterogeneities. Further, the lifetime of dynamic
heterogeneities can be measured when $F_4(t_{m2})$ for
$t_{m1}\!=\!t_{m3}\!=\!\mathrm{const.}$ is recorded. In this
experiment, identical filters are applied to probe the dynamic states
of an ion during $t_{m1}$ and a time $t_{m2}$ later during $t_{m3}$
so that an exchange of the dynamic state results in a decay of
$F_4(t_{m2})$, cf.\ below. Here, we will demonstrate that although
the lattice periodicity leads to imperfect dynamic filters in
applications on crystalline ion conductors $F_3(t_{m3})$ and
$F_4(t_{m2})$ still yield valuable insights into the nature of silver
dynamics in $\mathrm{Ag_7P_3S_{11}}$.

\begin{figure}
\includegraphics[angle=0,width=8.5cm]{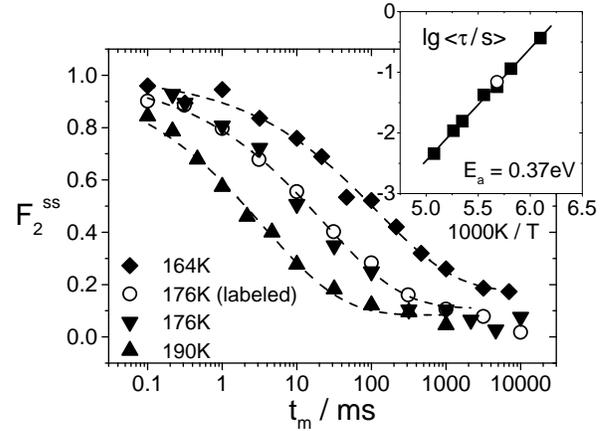}
\caption{$F_2^{ss}(t_m)$ for polycrystalline $\mathrm{Ag_7P_3S_{11}}$
at various temperatures ($t_p\!=\!100\mu s$). We compare results for
samples containing 48\% $^{109}$Ag (solid symbols, Ref.\
\cite{Vogel_JNCS}) and 97\% $^{109}$Ag (open symbols), respectively.
The dashed lines are interpolations with a modified KWW function.
Inset: Mean correlation times $<\!\tau\!>$ together with an Arrhenius
interpolation. }
\end{figure}

In Fig.\ 1, we show $^{109}$Ag NMR 2T-CF for $\mathrm{Ag_7P_3S_{11}}$
at selected temperatures. The silver jumps lead to temperature
dependent, non-exponential decays of $F_2(t_m)$. Experiments for the
labelled sample and for a sample with natural isotopic abundance
\cite{Vogel_JNCS} yield identical results. Fitting to a modified
Kohlrausch-Williams-Watts (KWW) function \cite{KWW},
$(1-C)\exp[-(t/\tau)^{\beta}]\!+C$, the non-exponentiality is
reflected by a stretching parameter of $\beta\!=\!0.42$. The plateau
value $C\!\approx\!0.12\!\approx\!1/8$ indicates eight occupied,
magnetically distinguishable silver sites \cite{Vogel_JNCS}. The mean
correlation time can be calculated according to
$<\!\tau\!>\,=\!(\tau/\beta)\,\Gamma(1/\beta)$ where $\Gamma(x)$ is
the $\Gamma$ function. Its temperature dependence is well described
by an Arrhenius law with activation energy
$E_{a}\!=\!0.37\mathrm{eV}$, see inset. We carefully checked that
neither spin-lattice relaxation ($T_1\!\approx\!30\mathrm{s}$) nor
spin diffusion affect the data for mixing times shorter than about a
few seconds.

The origin of the non-exponential silver dynamics in
$\mathrm{Ag_7P_3S_{11}}$ is revealed by $F_3(t_{m3})$. It is shown in
Fig.\ 2 for the dynamic filters
$t_{m1}\!=\!10\mathrm{ms},40\mathrm{ms}$ and $T\!=\!176\mathrm K$. In
particular for the longer $t_{m1}$, $F_3(t_{m3})$ decays much more
slowly than $F_2(t_{m3})$. In studies of supercooled liquids, the
theoretical expectations for purely heterogeneous dynamics,
$F_3^{het}(t_{m3})$, and purely homogeneous dynamics,
$F_3^{hom}(t_{m3})$, can be expressed in a model-free way based on
the measured 2T-CF \cite{3T,TrachtDiss,HeuerOkun}. For
$\mathrm{Ag_7P_3S_{11}}$, the presence of a small number of discrete
NMR frequencies, which are discontinuously adopted, complicates the
analysis. In general, one has to specify the crossover paths between
the different sites/$\omega$ to define the purely heterogeneous and
the purely homogeneous scenario. Since detailed information is not
available, we restrict ourselves to the analysis of two extreme
cases. For reasons of simplicity, we assume that there are eight
distinguishable sites on a cubic lattice. While the frequencies of
every other site in each dimension are identical in the ordered (O-)
model, the $N\!\approx\!8$ values of $\omega$ are randomly
distributed in the disordered (D-) model.

\begin{figure}
\includegraphics[angle=0,width=6cm]{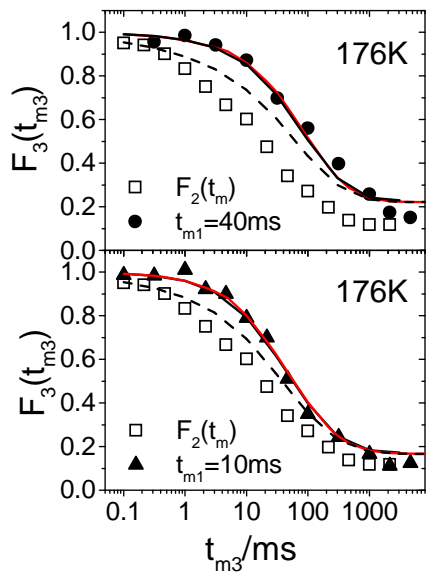}
\caption{$F_3(t_{m3};t_{m1})$ for $\mathrm{Ag_7P_3S_{11}}$ at
$T\!=\!176\mathrm K$ ($t_p\!=\!200\mu s$). The results for
$t_{m1}\!=\!10\mathrm{ms},\,40\mathrm{ms}$ are compared with
$F_3(t_{m3};t_{m1}\!\rightarrow\!0)\!\approx\!F_2^{cc}(t_{m3})\!\approx\!F_2(t_{m3})$.
The respective expectations for purely homogeneous dynamics (dashed
lines) and purely heterogeneous dynamics (solid lines) are included.
The expectations for heterogeneous dynamics in the O-model (black)
and the D-model (gray) are nearly identical.}
\end{figure}
For the following calculations, it is useful to rewrite
$F_3\!\propto\!1/2\{\langle\,\cos[(\omega_1\!-\!\omega_4)t_{p\,}]\,\rangle+
\langle\,\cos[(\omega_1\!+\!\omega_4\!-\!2\omega_2)t_{p\,}]\,\rangle\}$.
Since $\Delta \omega t_p\!\gg\!2\pi$, ions showing
$\omega_1\!=\!\omega_4$ and $\omega_1\!=\!\omega_2\!=\!\omega_4$
contribute to the first and the second term, respectively. We denote
$p^k(t)$ as the probability that after a time $t$ an ion occupies a
site with identical NMR frequency. If dynamic heterogeneities exist,
this probability depends on the jump rate as indicated by the index
$k$. In general, we can write
\begin{eqnarray}
&F_2(t_m) =\langle p^k(t_m) \rangle \:\:\mathrm{and}\:\:&\label{f2gen}\\
\!&\!F_3(t_{m1}\!,t_{m3})\!=\!1/2 \,\langle p^k(t_{m1}\! +\!
t_{m3})\! +\! p^k(t_{m1}) p^k(t_{m3}) \rangle.\:\:&\label{f3gen}
\end{eqnarray}
Here, the brackets denote the average over the rate distribution
$G(k)$ that is present in the case of dynamic heterogeneities, but
becomes a delta-function for purely homogeneous dynamics.

The homogeneous limit can be discussed in a model-free way. Since
a rate distribution is absent, one directly has
$p^k(t)\!=\!F_2(t)$ and, thus,
\begin{equation}
F_3^{hom} (t_{m1}\!,t_{m3})=  (1/2) [F_2(t_{m1}\!+\! t_{m3}) +
F_2(t_{m1}) F_2(t_{m3})].
\end{equation}
For the heterogeneous limit, we first discuss the O-model. In this
case, one can specify $p^k(t)$ from simple arguments. For a jump
rate $k$ and one-dimensional dynamics, the probability to be at a
site with the same NMR frequency is $(1/2)[1\!+\!
\exp(-2kt)]$\cite{3T}. Since the dynamics along the different
dimensions are independent of each other $p^k(t) = \{(1/2)[1\!+
\!\exp(-2kt)]\}^3$ follows. Using this expression for $p^k(t)$ one
can determine the distribution $G(k)$ from Eq.\ \ref{f2gen} and
use it to calculate $F_3^{het,O}$ via Eq.\ \ref{f3gen}. Now, we
turn to the heterogeneous scenario of the D-model. For a single
jump rate $k$, one has
\begin{equation}\label{F2_D}
p^{\,k}(t_m)=\frac{1}{N}\!+\!\left(1\!-\!\frac{1}{N}\right)\exp(-kt_m).
\end{equation}
With this choice of $p^{\,k}(t_m)$, Eq.\ \ref{f3gen} can be rewritten
as  \begin{eqnarray}\label{HETRJ}
F_{3}^{het,D}(t_{m1},t_{m3})=&F_2(t_{m1}\!+t_{m3})+
\frac{1}{2N}[F_2(t_{m1})+\nonumber\\&F_2
(t_{m3})-F_2(t_{m1}\!+t_{m3})-1].
\end{eqnarray}
Actually, we use $N\!=\!8.4$ for the D-model where this value results
from a fit of Eq.\ \ref{f2gen} to the experimental data.

In Fig.\ 2, we see that the experimental data for
$t_{m1}\!=\!10\mathrm{ms},40\mathrm{ms}$ agree well with
$F_3^{het}(t_{m3})$ obtained within both the O-model and the D-model.
The minor deviations in the plateau regime ($ t_{m3}\!>\! 1s $) are
likely due to an onset of spin diffusion. In contrast,
$F_3^{hom}(t_{m3})$ distinctly deviates from the experimental
results. The agreement of both limiting cases with respect to the
heterogeneous limit clearly shows that, as is the case for
supercooled liquids, the heterogeneous limit can be characterized in
a basically model-free manner. We conclude that correlated
back-and-forth jumps on the timescale of the silver relaxation are
not a relevant feature of the dynamics.

Finally, the lifetime of the dynamic heterogeneities in
$\mathrm{Ag_7P_3S_{11}}$ is measured by recording $F_4(t_{m2})$ for
$t_{m1}\!=\!t_{m3}\!=\!t_s\!\approx\,<\!\tau\!>$. If there were
perfect dynamic filters, slow silver ions ($\tau\!>\!t_s$) would be
selected during $t_{m1}$ and a time $t_{m2}$ later it would be
checked whether these ions are in the same dynamical state during
$t_{m3}$, cf.\ Eq.\ \ref{F4}. Thus, $F_4(t_{m2})$ would decrease when
initially slow ions become fast ($\tau\!<\!t_s$) during $t_{m2}$
until the re-equilibration of the respective subensemble is complete.
In our case, the signal results from all ions showing
$\omega_1\!=\!\omega_2$ and $\omega_3\!=\!\omega_4$ so that fast ions
that occupy periodic sites at the relevant times contribute, too.
Detailed random-walk simulations based on the O- and the D-model have
shown, however, that despite the imperfection of the dynamic filter
$F_4(t_{m2})$ is well suited to measure the timescale of exchange
processes between fast and slow ions.

\begin{figure}
\includegraphics[angle=0,width=7cm]{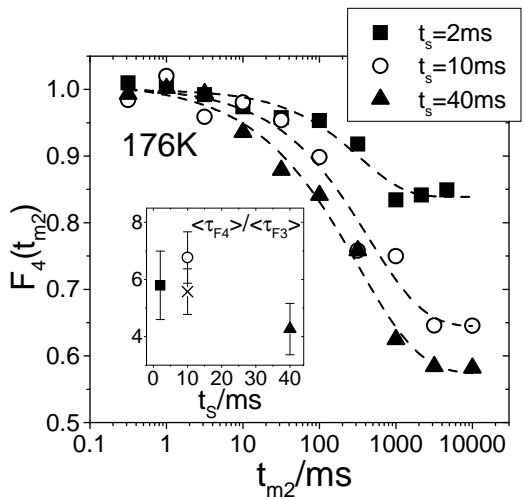}
\caption{$F_4(t_s,t_{m2},t_s)$ for $\mathrm{Ag_7P_3S_{11}}$ at
$T\!=\!176\mathrm K$ ($t_p\!=\!200\mu s$). The filter times $t_s$ 
are indicated. Lines:
Interpolations with a modified KWW function. The inset shows the
scaled mean decay times $<\!\tau_{F4}\!>\!/\!<\!\tau_{F3}\!>$. For
comparison, one data point for $T\!=\!182\mathrm K$ is included (X).
For $t_s = 2ms$ and for $T\!=\!182\mathrm K$ $<\!\tau_{F3}\!>$ has
been estimated from $F_3^{het}$, as calculated using Eq.\
\ref{HETRJ}.}
\end{figure}

The curves $F_4(t_{m2})$ for $\mathrm{Ag_7P_3S_{11}}$ at
$T\!=\!176\mathrm K$ and various $t_s$ are shown in Fig.\ 3.
Non-exponential decays to different plateau values for
$t_{m2}\!\rightarrow\!\infty$ indicate exchange processes between
slow and fast silver ions until the re-equilibration of the
respective subensemble is complete. To quantify the exchange rate the
decays are fitted to a modified KWW function. We obtain mean time
constants $<\!\tau_{F4}\!>\,=\!390\!-\!660\mathrm{ms}$ that do not
significantly depend on the filter time $t_s$. These values are on
the same order as $<\!\tau_{F3}\!>$ and, hence, the exchange
processes between slow and fast ions of the distribution occur on the
timescale of the slow silver jumps. However, as indicated by
$<\!\tau_{F4}\!>\!/\!<\!\tau_{F3}\!>\;\approx\!6$, see inset, it
takes a few rather than a single jump relaxation process to loose
memory about the mobility, suggesting that high and low energy
barriers are not randomly distributed in the unit cell of 
$\mathrm{Ag_7P_3S_{11}}$.

In summary, the observation of NMR MT-CF provides valuable insights
into the mechanism of ion motion, not available by other experimental
methods. The silver jumps in the crystalline ion conductor
$\mathrm{Ag_7P_3S_{11}}$ are governed by a broad rate distribution
within which exchange processes between slow and fast ions take place
on the timescale of the slow jumps. Further, the purely heterogeneous
scenario applies to the relaxation of the non-fast silver ions and,
hence, back-and-forth jumps are not relevant on the time scale of the
depopulation of the silver sites. At first sight, this seems to
contradict the presence of back-and-forth motions inferred from the
frequency dependent electric conductivity $\sigma(\nu)$. Both results
might be reconciled if one assumes that only very fast ions
contribute to the dispersion of $\sigma(\nu)$, whereas ionic motion
on the typical timescale of the silver relaxation is random-walk
like. This conclusion is backed up by MD simulations on glassy ion
conductors where the back-jump probability was found to depend
strongly on the waiting times at the sites \cite{Lammert,Vogel_MD}.

Funding of the Deutsche Forschungsgemeinschaft (DFG) through the
Sonderforschungsbereich 458 is gratefully acknowledged. M.\ V.\
thanks the DFG for funding through the Emmy Noether-Programm.

\end{document}